\documentclass[aps, prd, longbibliography, reprint]{revtex4-1}

\usepackage{graphicx}
\usepackage{hyperref}
\usepackage{upgreek}
\usepackage{wasysym}   % diameter&arrow symbols.

\begin{document}

\title{Hard bremsstrahlung from a high-voltage atmospheric discharge and its anisotropy}

\author{A.V.~Agafonov}
\affiliation{P.N.~Lebedev Physical Institute of the Russian Academy of Sciences (FIAN), Moscow, 119991, Leninsky pr., 53}
\author{A.V.~Oginov}
\email{oginov@lebedev.ru}
\affiliation{P.N.~Lebedev Physical Institute of the Russian Academy of Sciences (FIAN), Moscow, 119991, Leninsky pr., 53}
\author{A.A.~Rodionov}
\affiliation{P.N.~Lebedev Physical Institute of the Russian Academy of Sciences (FIAN), Moscow, 119991, Leninsky pr., 53}
\author{V.A.~Ryabov}
\affiliation{P.N.~Lebedev Physical Institute of the Russian Academy of Sciences (FIAN), Moscow, 119991, Leninsky pr., 53}
\author{V.A.~Chechin}
\affiliation{P.N.~Lebedev Physical Institute of the Russian Academy of Sciences (FIAN), Moscow, 119991, Leninsky pr., 53}
\author{K.V.~Shpakov}
\affiliation{P.N.~Lebedev Physical Institute of the Russian Academy of Sciences (FIAN), Moscow, 119991, Leninsky pr., 53}

\date{\today}

\begin{abstract}

The results of the experiments on recording hard gamma radiation and measurements of its angular distribution at the initial stage of a laboratory high-voltage atmospheric discharge are presented. The experiments were performed on an ERG installation at a voltage of $\sim 1$~MV, an atmospheric discharge current of up to 12~kA, and a gap of 0.55~m. The duration of the voltage pulse was about 1~$\upmu$s with a pulse rise time of 150--200~ns. The radiation was recorded by an assembly of 10 identical scintillation detectors installed each 10$^\circ$ around the circumference of a quarter of a circle with a curvature of 1~m. In order to separate the radiation with energies from 20~keV to 1.5~MeV, Al and Pb filters of different thicknesses were used.
The obtained results show that, as a rule, a multi-beam radiation pattern and several bursts of radiation (each with a directional pattern) are recorded in each shot. In a considerable number of ``shots'', hard radiation with photon energies comparable to or exceeding the maximum electron energy corresponding to the applied voltage is recorded. In these cases, a needle-like radiation pattern is observed, including at large angles to the axis of the discharge. This may indicate the acceleration of electrons in different plasma channels.

\end{abstract}


\maketitle

\section{Introduction}

Experimental and theoretical studies of radiation (high-energy electrons, positrons, radio and gamma radiation, neutrons) and the processes accompanying its generation during the development of an atmospheric discharge have been conducted for many years and constitute one of the rapidly expanding fields of physics. These efforts are aimed at elucidating the generation mechanisms of high-energy penetrating radiation in electrical discharges in the atmosphere, as well as at establishing the connection between these mechanisms and the processes of the initiation and development of lightning. A detailed review of the phenomena observed in this research area, referred to as ``high-energy atmospheric physics'', is given in~\cite{1}.

New data on the characteristics of thunderstorm processes obtained during the past 10--15 years show that the processes occurring in a thundercloud are much more complicated than previously thought. The experiments in the thunderstorm atmosphere are conducted both on ground installations, and from aircraft and balloons~\cite{2, 3, 4, 5, 6, 7, 8, 9}. According to the long-term investigations of lightning events, which were carried out during the recent decades at the height of the Tien Shan mountain cosmic ray station, it was demonstrated that gamma-rays, electrons and neutrons can play a key role in the initiation and development of the atmospheric discharge~\cite{3, 10, 11, 12, 13, 14}. Similar results have been obtained in Aragats~\cite{15, 16} and at other sites, both high-altitude~\cite{17} and sea-level~\cite{18, 19} ones.

The complexity of full-scale measurements and insufficient knowledge of the physical processes behind the discharge formation dictate the need for the research using laboratory facilities. The results of this research also indicate the presence of penetrating radiation during electrical discharges~\cite{20, 21, 22, 23, 24, 25, 26}.

In order to study the laboratory analog of a high-altitude atmospheric discharge, an ERG installation was designed and built at the Lebedev Physical Institute. A series of experimental studies were carried out to investigate the breakdown of runaway electrons and the generation of radiation at the initial stage of the atmospheric discharge~\cite{27, 28, 29, 30, 31}. In our work~\cite{31}, attention was drawn to the fact that the signals recorded by scintillation detectors, even those placed behind a thick (10~cm Pb) shield, can be caused not only by neutrons, but also by hard gamma radiation. However, the energies of the gamma quanta must greatly exceed the voltage applied to the discharge gap. Such anomalous excess was to a greater or lesser degree observed in~\cite{32, 33, 34}, whereas in~\cite{34}, in contrast to~\cite{32, 33}, the fraction of the runaway electrons with anomalous energies (or gamma quanta) was small. In most experiments at laboratory facilities the measured energies of the gamma quanta did not exceed 100--300~keV~\cite{20, 21, 22, 23, 24, 25, 26}. Although it was noted in~\cite{21} that some of the hard radiation signals corresponded to the gamma quanta with energies of several megaelectronvolts, it was assumed that a superposition of the signals associated with the x-ray quanta of lower energies occurred, with most of these quanta having energies of about 100~keV.

The anisotropy of gamma radiation in a laboratory atmospheric discharge can arise for a number of reasons. If the bremsstrahlung is produced by a stream of relativistic electrons, then it must exhibit a pronounced anisotropic character. Correspondingly, detection of strong anisotropy (needle-like radiation) could be the evidence for the existence of relativistic electrons. In this case, the angle within which the radiation is detected would allow us to determine the approximate energy of the electrons. However, the form of discharge, which can be very bizarre, never repeats not only ``from shot to shot'', but even in a series of dozens of shots.

Radiation occurs mainly at the initial (streamer-leader) stage of the discharge before the cathode and anode coronas moving towards each other join to form a discharge channel (channels) against the background of a sharply increasing discharge current. Rarely recorded are the radiation flashes, which can be timed to the closure of the opposing coronas. It was noted in~\cite{25} that the most severe radiation occurs at this instant. At the initial stage of the discharge a lot of streamer-leader channels having different directions are, as a rule, developed both at the cathode and at the anode. In which direction the discharge will develop depends on the ``surviving'' competitive channels.

Braking radiation is created by ``runaway'' electrons. At relatively low energies of the runaway electrons, the radiation in the forward hemisphere does not differ much from the isotropic one. However, the detectors can record clear radiation anisotropy associated with the direction of the channel evolution (along the axis of the discharge gap or sideways therefrom). In particular, this can happen due to lower intensity of the radiation incident on the detectors located at relatively large distances and at large angles to the direction of the channel evolution. Moreover, additional electron acceleration due to collective effects in the plasma is possible in the generated channels, and this effect is not localized in the near-cathode region.

Thus, even in the case of such a relatively low-energy bremsstrahlung, its angular distribution can be of a ``directional'' character and can be obtained within a small solid angle. Therefore, recording a strong anisotropy of the angular distribution of the radiation does not allow us to conclude with sufficient confidence that the radiation is produced by the high-energy electrons. In the simplest case, the anisotropy of high-energy radiation can be recorded by using filters of different thicknesses. Transmission or absorption of the radiation in these filters provide a means to obtain a fairly accurate estimate of the energy of the gamma quanta.

It should be noted that, in our case, an additional agent affecting the direction of the channel evolution (direction of electron ``runaway'') is the annular anode (the outer cylindrical electrode of the forming line of a former high-current electron accelerator, 2~m in diameter (see Fig.~\ref{fig:1}), which gives rise to a sufficiently high radial component of the electric field in the near-cathode region.

\section{Experimental setup}

\begin{figure}
\resizebox{0.45\textwidth}{!}{
\includegraphics*{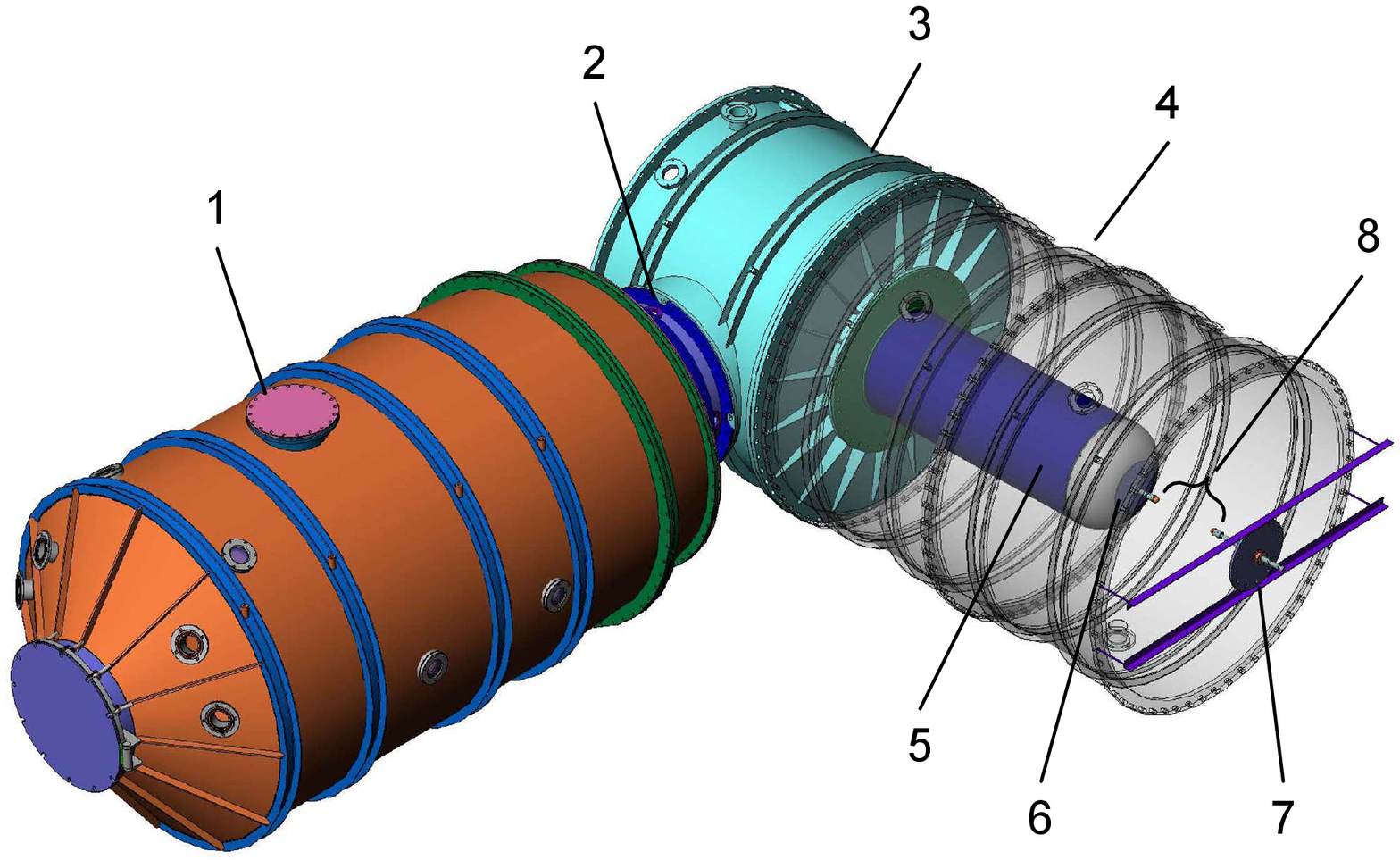}}
\caption{ \label{fig:1} Scheme of experimental setup: \textit{1}---oil-filled GIN, \textit{2}---oil-filled coupling section, \textit{3}---oil-filled section of high-voltage insulator, \textit{4}---shells of air section (shown transparent), \textit{5}---oil-air bushing insulator, \textit{6}---cathode electrode system, \textit{7}---anode electrode system, \textit{8}---main discharge gap.
}
\end{figure}

The experiments were performed on an ERG high-current accelerator reconstructed to study the high-voltage atmospheric discharge (Fig.~\ref{fig:1}). The total duration of the voltage pulse is varied depending on the selected discharge gap length ranging from 0.35 to 1~$\upmu$s with the rise time of the voltage pulse front of 150--200~ns, maximum voltage of 1.2~MV and discharge current of 10--15~kA.

A combination of the electrodes in the form of a needle cathode inside a cone and a hemispherical mesh anode was used in the experiments. This configuration was proved to be the most effective for generating radiation~\cite{30, 31}. The influence of the shape of the electrodes on the generation of radiation in a laboratory atmospheric discharge was also mentioned in~\cite{35}.

In order to record the radiation in an atmospheric discharge with temporal resolution at the level of several nanoseconds, a FEU-30 high-current pulse photomuliplier with a sensitivity of 1000--5000~A/lm was used; its input window was coupled to fast scintillators. The scintillators were made of polystyrene incorporating scintillating additives (p-terphenyl + POPOP) and were matched to the sensitivity of the photomultiplier in the spectral range of 360--440~nm.

All detectors were calibrated using a RINA standard x-ray source~\cite{36}. For each of them, a correction factor was determined and then taken into account when processing the recorded signals.

\begin{figure}
\resizebox{0.45\textwidth}{!}{\includegraphics*{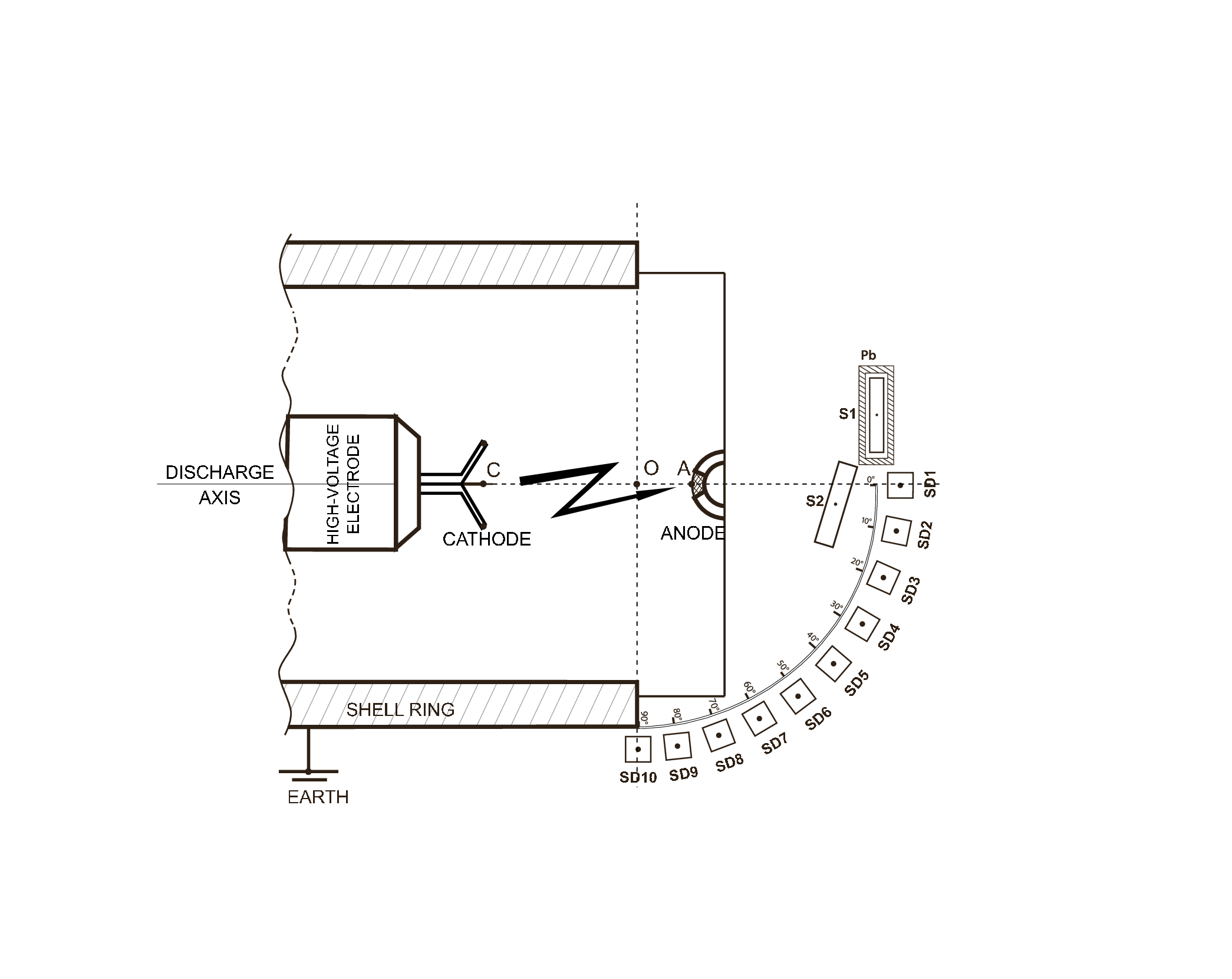}}

\caption{Arrangement of detectors.} \label{fig:2}
\end{figure}

To measure the angular distributions of the radiation, a special system involving 10 scintillation detectors (SD1--SD10) installed each 10$^\circ$ around the circumference of a quarter of a circle with a curvature of 1~m was created (Fig.~\ref{fig:2}). We further refer to these detectors as the ``on-arc'' detectors. The center of the circle was on the discharge axis at a distance of about 2/3 of the discharge gap length from the cathode. All detectors were placed in the horizontal plane of the discharge axis; the discharge axis passed through the detector SD1. The detector SD10 was at the periphery of the arc at an angle of 90$^\circ$ to the axis of the discharge.

All on-arc detectors were shielded with the filters of equal thickness. To isolate the radiation with photon energies higher than 20~keV, a constructive 3-mm-thick Al filter (detector housing) was used. To isolate the radiation with photon energies higher than 100~keV a 3-mm-thick Pb filter was used, and a 10-mm-thick Pb filter was used to isolate the photons with energies above 350~keV. Finally, for a hypothetical case of generation of hard radiation with photon energies comparable to that acquired by electrons at the maximum voltage, a 50-mm-thick Pb filter (providing tenfold attenuation of the gamma radiation having an energy of 1.5~MeV) was used~\cite{37}.

In addition to the arc assembly, two more detectors with rectangular scintillators, 150$\times$150$\times$50~mm in size, were used: a high-threshold forward detector S1 and a low-threshold forward detector S2  (Fig.~\ref{fig:2}). The detector S1 was placed in a 50-mm-thick Pb shielding and recorded the hard radiation component. The detector S2 was covered with a thin ($\sim 10$~$\upmu$m) layer of an aluminum (Al) foil and opaque paper. It recorded the discharge radiation with a minimum threshold energy of 10~keV. The arrangement of the detectors is shown in Fig.~\ref{fig:2}.

\section{Estimates for bremsstrahlung from a high-voltage discharge in air}

We assume that at the initial stage of the atmospheric discharge the bremsstrahlung is formed by a beam of runaway electrons. Let us consider the characteristic angular distributions of the bremsstrahlung produced by the electron beam at the initial stage of a discharge in air at atmospheric pressure in terms of the simplest model.

The beam parameters taken for the estimates are the following: current 5~mA, duration 100~ns, total number of electrons $\approx 3\times10^9$. We will characterize the beam by various energy distribution functions, and assume that it propagates along the axis of the discharge gap. We will also assume that the electron energy is constant, i.e. the radiation losses are compensated by the acceleration in the external field. We also neglect multiple scattering of electrons. Therefore, in fact, we sum the radiation of individual electrons incident on the detector that have different energies; the number of the electrons within different energy ranges is given by the distribution function. In addition, we will assume that the signal from the detectors is proportional to the total energy $Q$ of the quanta incident on the detector rather than to their number.

In the calculations, we adopt the real geometry of the experiment with ten detectors placed each 10$^\circ$ around the circumference of a quarter of a circle. Therefore, the dependences for the total energy of the quanta are calculated for the specific location of the corresponding detectors, i.e. pointwise.

The bremsstrahlung cross section is estimated in the Born approximation, when the initial $v_0$ and final $v$ electron velocities are sufficiently large: $Z/137\approx0.05\ll v_0/c \approx \sqrt{2W_0/(mc^2)}$. Here $Z=7$ is the ordinal number of the air nuclei, $m$ is the electron mass, $W_0$ is the initial kinetic energy of the electron. The initial electron energy is assumed to exceed several kiloelectronvolts. Naturally, the final kinetic energy of the electron $W_{\max}$ is also to satisfy this condition; otherwise, the Born approximation is inapplicable. Thus, we can consider both the nonrelativistic case, for example, $W_0=10$~keV, and the relativistic case, for example, $W_0=10000$~keV.

Calculations were performed in the MathLab environment using the formula for the bremsstrahlung differential cross section for a quantum emitted within solid angle with energy $W$ ${\mathrm d}\Omega = d \cos \theta {\mathrm d} \varphi$, given in~\cite{38}.

The calculations were carried out for several types of distribution function:

1) the Maxwellian energy distribution function $W$ with a sufficiently large ``temperature'' $T_\mathrm{M}$, truncated at the energy $W_{\max}$:

$$ f(W)= \exp (-W/T_\mathrm{M}) \sqrt{W/T_\mathrm{M}};$$ 

2) the distribution functions of runaway electrons in the atmospheric discharge from Refs.~\cite{39, 40, 41} with the energy spectrum of the runaway electrons in the high-energy region given as $f \sim W^{-\beta}$; however, the data on the exponent differ from paper to paper. Thus, $\beta \approx 0.8$ in~\cite{39}, and in a more recent paper~\cite{40} $\beta \approx 1.2$. Then, in~\cite{41} $\beta \approx 1$ is given with an additional peak in the region of maximum energies, is indicated.

Qualitative results appeared to be similar for all considered distribution functions.

\begin{figure}
\resizebox{0.45\textwidth}{!}{\includegraphics*{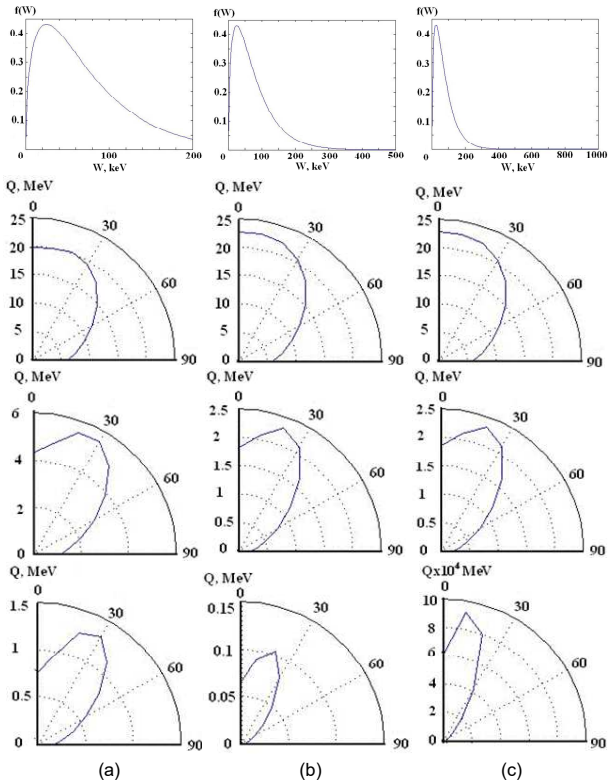}}
\caption{Maxwellian distribution functions with $T_\mathrm{M} = 50$~keV and $W_{\max} = 200$~keV (\textit{a}), 500~keV (\textit{b}), and 1000~keV (\textit{c}) (upper row). Vertically, under each distribution function, corresponding angular distributions of total energy $Q$ of quanta incident on detectors are given for different cut-off energies of quanta (quanta having energies below specified value are disregarded): $W_\mathrm{cut}= 10$~keV (second row); $W_\mathrm{cut}= 50$, 100 and 100~keV (third row); $W_\mathrm{cut} = 100$, 250 and 500~keV (bottom row).} \label{fig:3}
\end{figure}

For a Maxwellian distribution with $T_\mathrm{M} = 50$~keV and $W_{\max} = 200$~keV, 500~keV, and 1000~keV, the calculation results for the angular distributions of the total energy $Q$ of the quanta incident on the detectors placed at different angles are shown in Fig.~\ref{fig:3}; the quanta with energies below the specified value $W_\mathrm{cut}$ are disregarded. Cutting off the quanta with energies below the target value simulates the presence of the filters of different thicknesses covering the detectors. The calculated dependences are given for the actual geometry of the experiment; the total energy of the quanta $Q$ (in electronvolts) emitted at a given angle is averaged over the spectrum.

It can be seen from the above data that if the detectors are shielded with the filters cutting off only the quanta with energies below 10~keV (second row), the angular distributions are practically identical. That is, the contribution from the high-energy electrons is negligible, since their number is small (tails of the distribution functions). Nevertheless, the angular distributions have a pronounced anisotropy in the range from 0$^\circ$ to 90$^\circ$. However, in the near-axis region, they are practically isotropic within the range of 0$^\circ$--30$^\circ$. For thicker filters that cut off the quanta with energies lower than 50 and 100~keV (third row), a dip in the distributions appears near the axis; the distributions become anisotropic with a maximum at an angle of about 30$^\circ$. More importantly, the intensity of the radiation associated with hard quanta is an order of magnitude smaller than in the case of a thin filter transmitting all emitted photons having energies above 10~keV. In the case of thick filters that cut off quanta in an even larger region, with energies below 250~keV and 500~keV (fourth row, two right columns), the maximum of the distribution is shifted towards the axis. The position of the distribution peak depends on the maximum electron energy in the beam spectrum and shifts from 30$^\circ$ for a cut-off energy of 200~keV to $\sim 10^\circ$ for a cut-off energy of 1000~keV and thick filters. In addition to sharp narrowing of the angular distribution, the total energy of the quanta incident on the detector decreases by many orders of magnitude. This leads to a sharp drop in the detector signal and entails difficulties in its recording. Thus, for the case illustrated in the right-hand column of Fig.~\ref{fig:3}, the intensity of the radiation associated with hard quanta falls by 4 orders of magnitude. This actually means that, with fixed tuning of the scintillation detectors, the signals produced by the hard quanta  will, at best, be at the noise level.

\begin{figure}
\resizebox{0.45\textwidth}{!}{\includegraphics*{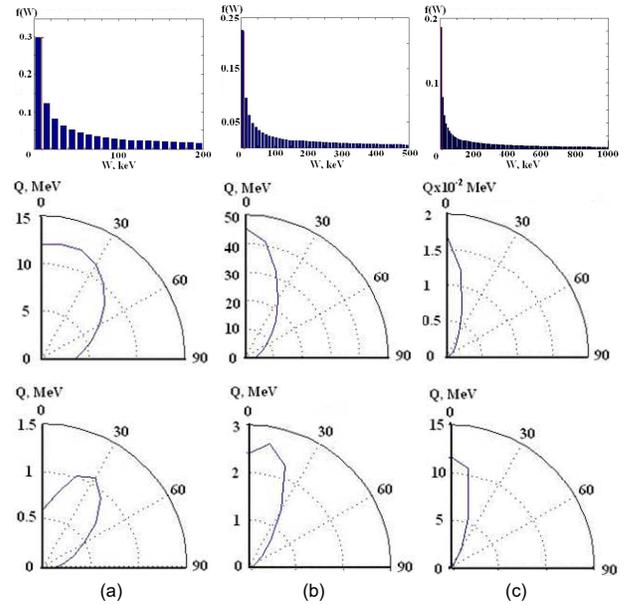}}
\caption{Power-law distribution functions with $\beta = 0.8$, $W_{\max} = 200$, 500 and 1000~keV (upper row), and corresponding \textit{a, b, c} calculated angular distributions of total energy $Q$ of quanta, photons with energies below specified values are cut off: $W_\mathrm{cut} =10$~keV (middle row); $W_\mathrm{cut} = 100$, 250, and 500~keV, respectively (bottom row).} \label{fig:4}
\end{figure}

For the runaway electron distribution functions of the form $f \sim W^{-\beta}$, reported in~\cite{30, 31, 32, 33, 34, 35, 36, 37, 38, 39, 40, 41}, the results of the calculations are similar. The angular distributions obtained for the maximum energies $W_{\max} = 200$, 500 and 1000~keV are shown in Fig.~\ref{fig:4}; the cut-off energy of the quanta is $W_\mathrm{cut}= 10$~keV and $\beta = 0.8$. The distribution functions (the upper row in Fig.~\ref{fig:4}) are given in a discrete form, since the solution of the kinetic equation describing the breakdown associated the with runaway electrons was obtained numerically.

The following circumstance draws attention. When the quanta below 10~keV are cut off, the total energy of the quanta incident on the detector is increased by an order of magnitude with increasing the maximum electron energy from 200~keV to 1~MeV (middle row). This is in contrast to the case of the Maxwellian distribution (Fig.~\ref{fig:3}), for which variation of the total energy was negligible due to a substantially smaller number of the high-energy electrons.

For the distribution functions describing the breakdown associated with runaway electrons in the case of a strong electric field, there is a peak at the ``tail'' of the distribution function in the region of maximum energies~\cite{41}. We take the value of the relative number of the electrons contained in the peak to be at the level of 10\% of the total number of electrons.

\begin{figure}
\resizebox{0.45\textwidth}{!}{\includegraphics*{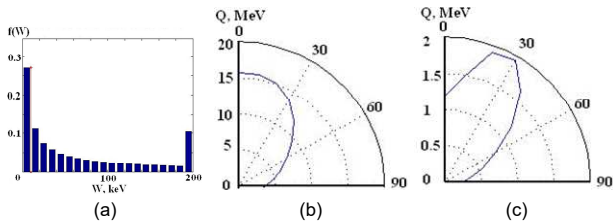}}
\caption{\textit{a}---Power-law distribution function with $\beta = 0.8$, $W_{\max} = 200$~keV with additional peak and calculated angular distributions of total energy $Q$ of quanta, photons with energies below specified values are cut off: \textit{b}---$W_\mathrm{cut}= 10$~keV, c---$W_\mathrm{cut} = 100$~keV.} \label{fig:5}
\end{figure}

At a small maximum energy $W_{\max} = 200$~keV and $\beta = 0.8$ (see Fig.~\ref{fig:5}), the results do not differ qualitatively from the case of the Maxwellian distribution (Fig.~\ref{fig:3}) and the distribution without the peak in the region of the maximum energies (Fig.~\ref{fig:4}).

However, for $\beta = 0.8$ as well as for large values of $\beta$, the presence of a peak in the distribution function in the region of maximum energies leads to the fact that, with increasing the maximum energy $W_{\max}$, the angular distribution becomes close to a needle-like shape with a maximum on the axis. Due to the decrease in the number of the low-energy quanta associated with the form of the distribution function and the presence of a peak in the region of maximum energies, the angular distribution has the shape similar to that shown in the right column of Fig.~\ref{fig:4}. The needle-like character of the angular distribution of the radiation with a maximum on the axis is conserved even if the high-energy quanta are cut off. This is in contrast to the Maxwellian distribution function, for which the radiation maximum appears at an angle to the axis rather on the axis itself (Fig.~\ref{fig:3}). 

Naturally, it is difficult to judge on the temporal structure of the radiation using the considered approach. However, a number of conclusions are obvious. If the detectors are shielded with thin filters that eliminate the low-energy quanta, then the linearly propagating beam radiation with different maximum electron energy reveals a weakly anisotropic character and the angular distributions are practically identical. As the thickness of the filters is increased and depending on the electron distribution function, a dip in the near-axis region appears. In this case, the angular distribution becomes narrower and, with increasing the maximum electron energy, the maximum of the distribution is angularly shifted towards the axis. In addition to the narrowing of the angular distribution, the total energy of the quanta incident on the detector changes by an order of magnitude. This corresponds to a drastic change in the amplitude of the detector signal and and entails difficulties in its recording. That is, with fixed tuning of the scintillation detectors, the amplitudes of the signals can vary by orders of magnitude.

\section{Measurement of bremsstrahlung angular distributions for a high-voltage atmospheric discharge}
\subsection{Results on the angular distributions of the radiation in the energy range above 20~keV}

In the corresponding experiments, the on-arc detectors were not covered by additional filters. In this case, 3-mm-thick Al detector housings acted as filters.

The obtained experimental data show that, with a single radiation pulse recorded by the low-threshold forward detector S2, a multi-lobe radiation pattern is observed. A similar picture arises even in the case when several pulses are recorded by the detector S2 in one shot. For each of these pulses, the on-arc detectors also record the signals, which appear as several flashes with their own radiation patterns.

\begin{figure}
\resizebox{0.45\textwidth}{!}{\includegraphics*{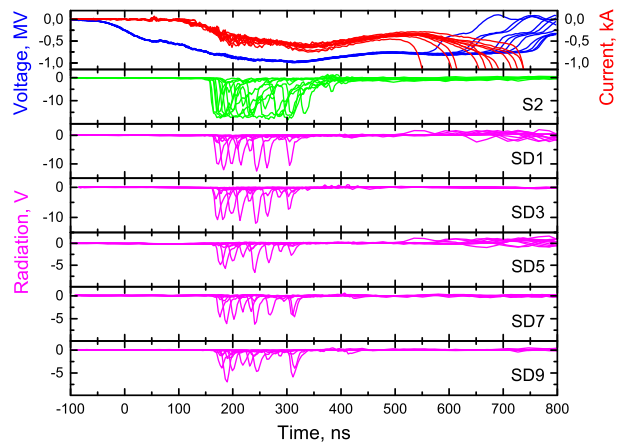}}
\caption{Superimposition of oscillograms obtained in 11~shots. Above are voltage and initial part (pre-pulse) of discharge current, below are oscillograms recorded by detector S2 and on-arc detectors SD1, SD3, SD5, SD7 and SD9.} \label{fig:6}
\end{figure}

Valuable information is provided by the superimposition of all shots in one of the series. Figure~\ref{fig:6} presents superimposed oscillograms of the radiation signals detected by the low-threshold forward detector S2 with a threshold of 10~keV and by several on-arc detectors with a threshold of 20~keV. The amplitude of the signals from the on-arc detectors falls by a factor of 2--3 with increasing the angle from 0$^\circ$ to 90$^\circ$ (the scale on the bottom three panels SD5, SD7 and SD9 is reduced). However, the amplitude remains large enough in the direction perpendicular to the axis. The 20~keV cut-off energy of the quanta is insufficient to resolve the structure of the discharge radiation.

The angular distributions of the radiation with photon energies above 20~keV bear little resemblance to the calculated ones, and are only somewhat similar to those shown in Figs.~\ref{fig:3}--\ref{fig:5}. This is due to the fact that the radiation is generated by the electrons accelerated in different directions, which create a large number of low-energy quanta, rather than by the electrons moving in one direction (along a straight line, from the cathode to the anode) in the form of an electron flow, as was assumed in the estimates.

\begin{figure}
\resizebox{0.45\textwidth}{!}{
\includegraphics*{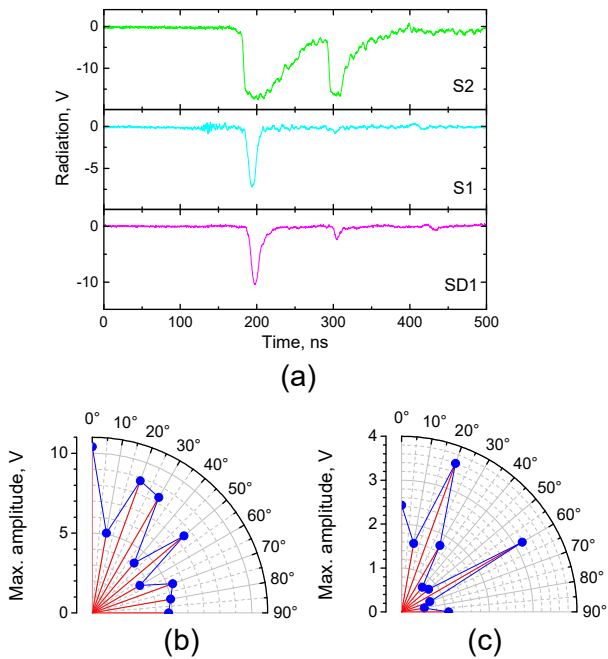}}
\caption{\textit{a}---Signals from low-threshold forward detector S2, high-threshold forward detector S1 and one of on-arc detectors (SD1); \textit{b}---angular distribution of radiation for two radiation flashes with photon energies above 20~keV (200~ns and 300~ns) recorded in the shot.} \label{fig:7}
\end{figure}

Figure~\ref{fig:7} shows the signals detected by the low-threshold forward detector S2, the high-threshold forward detector S1, the on-arc detector SD1, and the angular distributions of the radiation with photon energies above 20~keV recorded in one shot. It is seen that with increasing the hardness of the gamma quanta, the duration of the radiation pulses decreases. If the duration of the pulses of the total radiation with photon energies above 10~keV exceeds 100 and 50~ns, then for photon energies above 20~keV it is shorter than 20~ns, and is approximately twice as small for the gamma quanta with energies comparable to the ``applied voltage''.

In this and other similar series there are shots in which the on-arc detectors do not capture any signals, or the detector signals are by an order of magnitude lower than those in Fig.~\ref{fig:7}\textit{b}, for example. One can also mention the shots in which three pulses are recorded by the total radiation detector S2 with only one pulse recorded by the 90$^\circ$ arc detector.

\begin{figure}
\resizebox{0.45\textwidth}{!}{
\includegraphics*{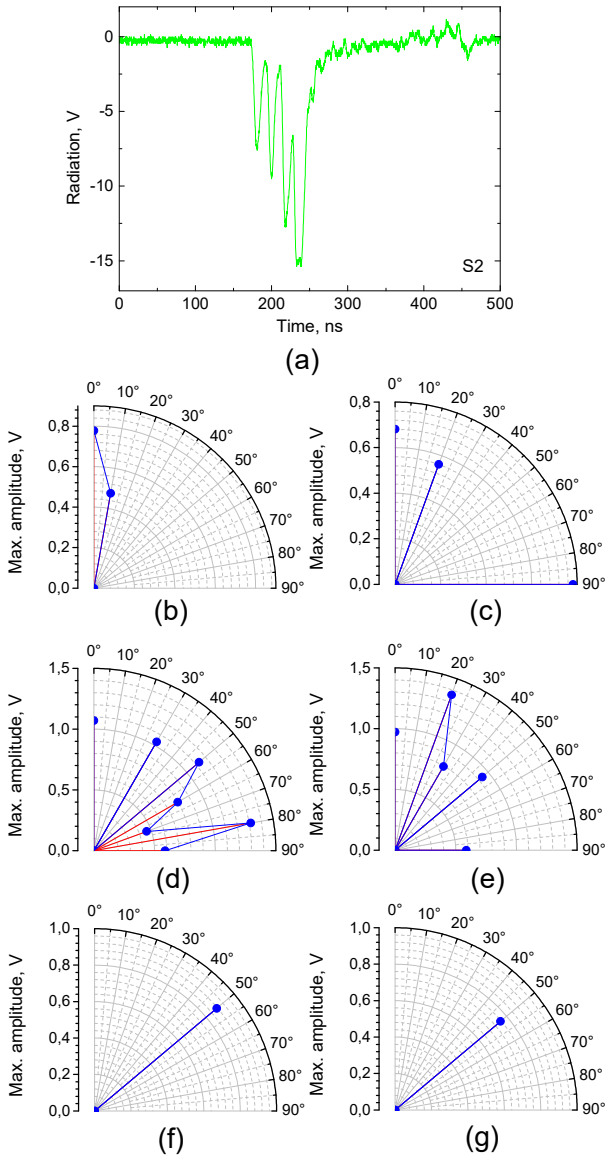}}
\caption{Signal recorded by low-threshold forward detector S2 \textit{a} and radiation angular distributions \textit{b--g} corresponding to peaks of recorded signal at 179~ns \textit{b}, 200~ns \textit{c}, 220~ns \textit{d}, 238~ns \textit{e}, 256~ns \textit{f} and 266~ns \textit{g}.} \label{fig:8}
\end{figure}

Thus, the development of the discharge occurs in different modes. Specifically, in Fig.~\ref{fig:7}, an intensive radiation with photon energies of above 20~keV is recorded continuously with respect to the angle. This indicates that electrons acquire a large energy. In another shot, shown in Fig.~\ref{fig:8}, the angular distribution appears as a set of ``needles'' with their amplitude being small but clearly defined against the background noise. In this case, only several of the ten on-arc detectors are triggered, and the number of the flashes in this shot does not coincide with that of the total radiation. This happens even though all six radiation flashes recorded by the on-arc detectors fall within the duration of the radiation recorded by the low-threshold forward detector S2, and the durations of these six flashes do not exceed 5--7~ns. The high-threshold forward detector S1 does not capture any signals.

The energy of the quanta on the order of 20~keV is the maximum energy accelerated electrons can acquire in this mode of the discharge formation. The stochastic character of the flashes and small amplitude of the signals indicate that the radiation of individual channels is detected. The electrons in this channels acquire the energy somewhat higher than 20~keV.

The information on the radiation angular distributions makes it possible (in some cases) to determine in which region of the discharge gap the radiation was formed. It can be assumed that triggering of the near-axis detectors (0$^\circ$--30$^\circ$) occurs due to the acceleration of electrons and appearance of the bremsstrahlung near the anode (cathode-directed streamers). We also suggest that the radiation directed at large angles to the discharge axis is probably generated in the near-cathode region by the electrons accelerated at an angle to the axis (anode-directed streamers). Naturally, such picture of radiation generation entails great difficulties in the calculations, especially since the energies of the photons accelerated from the near-cathode and from the near-anode regions can vary greatly, depending on the mode of the discharge development.

\begin{figure}
\resizebox{0.45\textwidth}{!}{
\includegraphics*{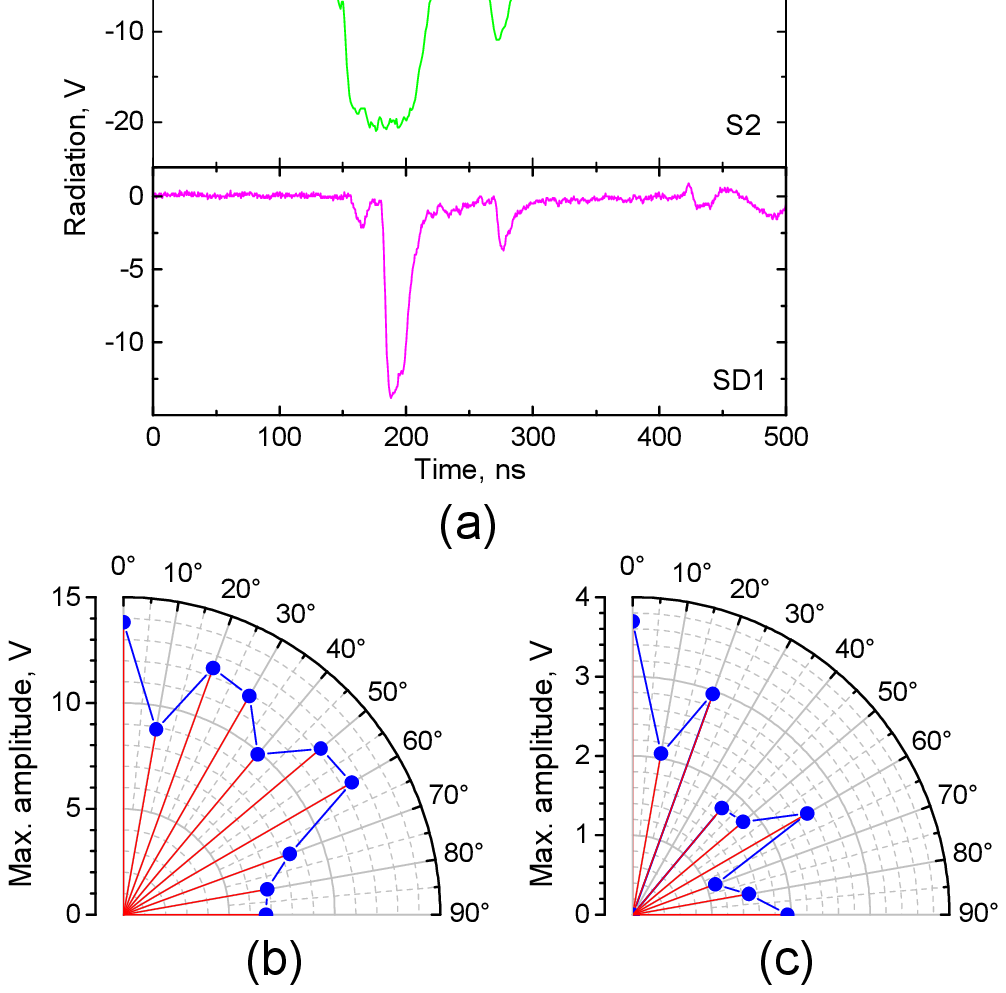}}
\caption{Signals from low-threshold forward detector S2, detector SD1 \textit{a} and radiation angular distributions \textit{b, c} corresponding to peaks of recorded signals at 200~ns \textit{b} and 285~ns \textit{c}.} \label{fig:9}
\end{figure}

If ``soft'' filters are used, quasi-isotropic angular distributions, similar to the calculated ones, are recorded mostly in intense radiation flashes. Figure~\ref{fig:9}\textit{b} shows an example of such distribution.

\subsection{Results on the angular distributions of the radiation in the energy range above 100~keV}

In the corresponding experiments, all on-arc detectors were covered with an additional 3~mm Pb filter, with tenfold attenuation of the gamma radiation having an energy of 100~keV (cut-off energy).

\begin{figure}
\resizebox{0.45\textwidth}{!}{\includegraphics*{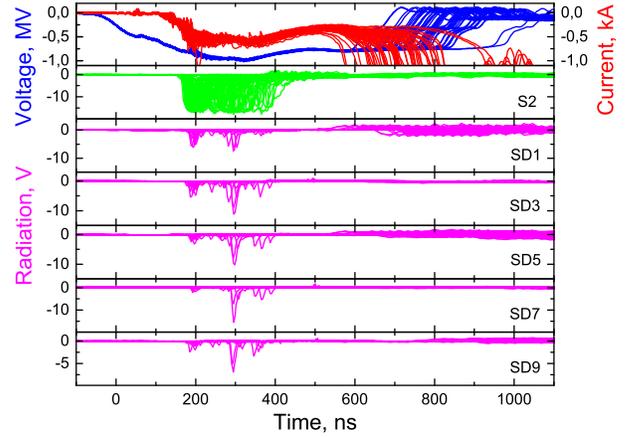}}
\caption{Superimposition of oscillograms obtained in 60~shots. Above are voltage and initial part (pre-pulse) of discharge current, below are oscillograms for  low-threshold forward detector S2, and for on-arc detectors SD1, SD3, SD5, SD7 and SD9.} \label{fig:10}
\end{figure}

Figure~\ref{fig:10} shows superimposed current and voltage signals (top panel) for 60~shots, the signals from radiation with photon energies above 10~keV (detector S3--total radiation), and the signals from radiation with energies above 100~keV (detectors SD1, SD3, SD5, SD7, SD9). In contrast to Fig.~\ref{fig:8}, all on-arc detectors indicate that there are three stages of the development of the discharge (with characteristic time of 90--100~ns), which are accompanied by the generation of the radiation with photon energies above 100~keV.

For the cut-off energy of 100~keV, the maximum amplitudes of the signals from the on-arc detectors are close to those obtained for the 20~keV cut-off energy (compare Fig.~\ref{fig:6} and Fig.~\ref{fig:10}). Intense flashes of radiation having rather continuous angular distributions or needle-like distributions can be recorded in one shot; combination of both can be recorded as well. As a rule, the amplitude of the intense flashes with a continuous angular distribution, exceeds the amplitudes of the needle-like radiation pulses by an order of magnitude or even higher than that.

\begin{figure}
\resizebox{0.45\textwidth}{!}{
\includegraphics*{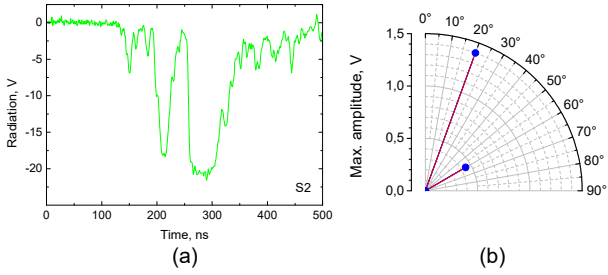}}
\caption{\textit{a}---oscillogram for low-threshold forward detector S2; \textit{b}---needle-like distribution of radiation with energies of quanta above 100~keV, corresponding to oscillogram peak at 285~ns.} \label{fig:11}
\end{figure}

An example of such needle-like distribution is shown in Fig.~\ref{fig:11}. The discharge develops at the maximum electron energies below 100~keV and only during the third radiation flash recorded by the low-threshold forward detector S2; two of the on-arc detectors detect flashes with photon energies above 100~keV.

\begin{figure}
\resizebox{0.45\textwidth}{!}{
\includegraphics*{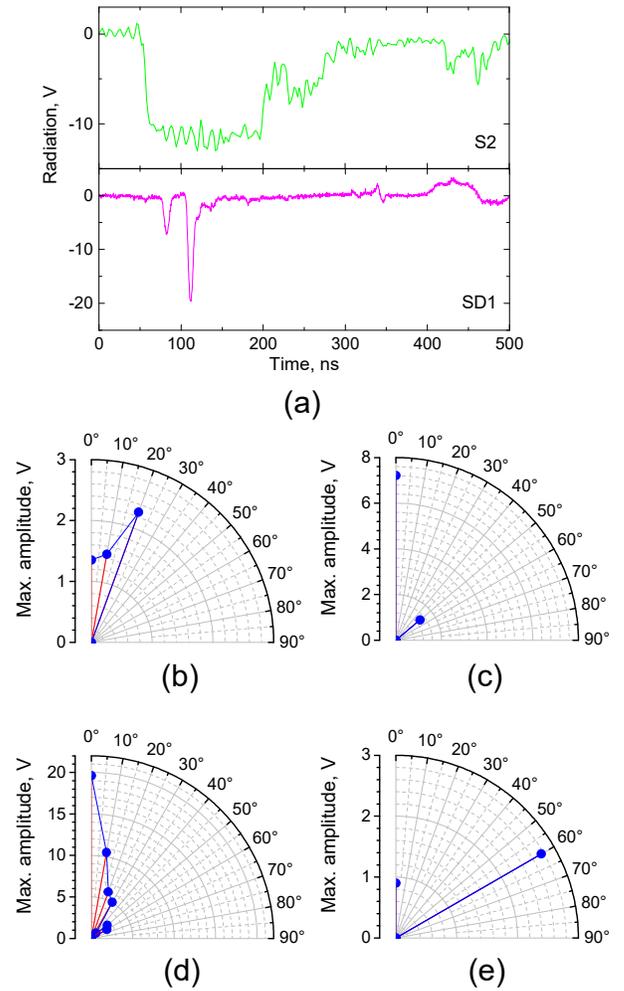}}
\caption{Above---oscillograms for low-threshold forward detector S2 and for on-arc detector SD1 \textit{a}, below---angular distributions of radiation with energies above 100~keV \textit{b--e}. Radiation flashes correspond to 60~ns, 80~ns, 115~ns and 225~ns.} \label{fig:12}
\end{figure}

Angular distributions of the radiation, which are close to the calculated ones, are also observed for high-intensity flashes. Figure~\ref{fig:12} shows the angular distributions of the radiation (5 flashes during the shot). The third distribution, which coincides in time with the middle of the unresolved radiation pulse from the detector S2, is in good agreement with the angular distribution calculated (see Fig.~\ref{fig:4}) for the power-law distribution function of the electrons.

\begin{figure}
\resizebox{0.45\textwidth}{!}{
\includegraphics*{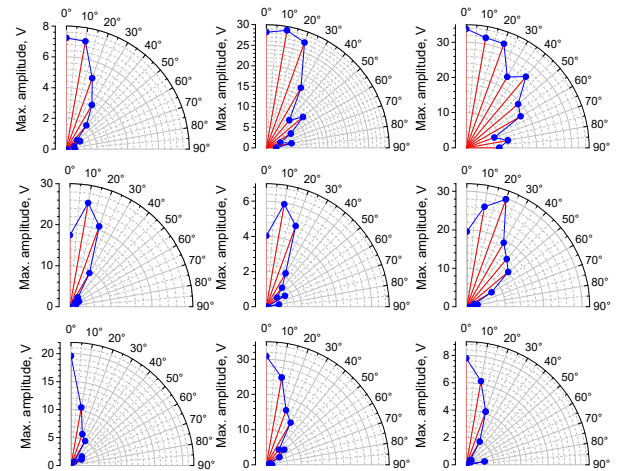}}
\caption{Examples of angular distributions close to calculated ones.} \label{fig:13}
\end{figure}

Below (Fig.~\ref{fig:13}) we provide several selected angular distributions, which were obtained in different shots and coincide well with those calculated for different conditions.

\subsection{Results on the angular distributions of the radiation in the energy range above 350~keV}

In the considered experiments, all on-arc detectors were covered with an additional 10-mm-thick Pb filter, with tenfold attenuation of the gamma radiation having an energy of 350~keV (cut-off energy).

\begin{figure}
\resizebox{0.45\textwidth}{!}{\includegraphics*{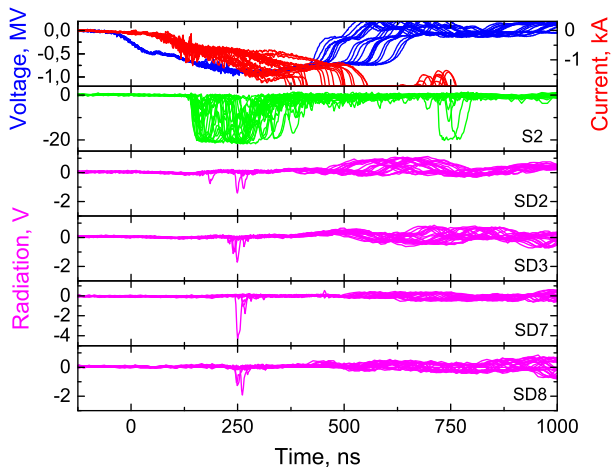}}
\caption{Superimposition of oscillograms in 24~shots. Upper panel presents voltage and pre-pulse of longitudinal discharge current; other panels present oscillograms for low-threshold forward detector S2, oscillograms for on-arc detectors SD2 (10$^\circ$), SD3 (20$^\circ$), SD7 (60$^\circ$) and SD8 (70$^\circ$).} \label{fig:14}
\end{figure}

Figure~\ref{fig:14} shows superimposed current and voltage signals (upper panel) for 24~shots, the signals from radiation with a photon energies above 10~keV (detector S2, total radiation), and the signals from radiation with energies above 350~keV (detectors SD1, SD2, SD3, SD7 and SD8).

\begin{figure}
\resizebox{0.45\textwidth}{!}{\includegraphics*{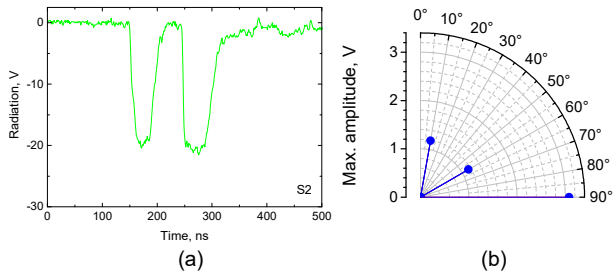}}
\caption{Oscillogram for low-threshold forward detector S2 (left) and needle-like angular distribution of radiation with photon energies above 350~keV (right) at 275~ns.} \label{fig:15}
\end{figure}

From the comparison of Fig.~\ref{fig:10} and Fig.~\ref{fig:14} it can be seen that the radiation with photon energies above 350~keV occurs mainly at the same instant (250~ns) as does the second radiation pulse with photon energies above 100~keV (Fig.~\ref{fig:10}). However, the angular distributions in this energy range, as a rule, reveal a needle-like character (see Fig.~\ref{fig:15}).

\begin{figure}
\resizebox{0.45\textwidth}{!}{\includegraphics*{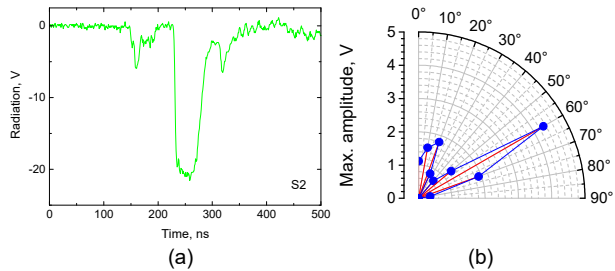}}
\caption{Oscillogram for low-threshold forward detector S2 (left) and piecewise-continuous angular distribution of radiation with photon energies above 350~keV (right) at 260~ns.} \label{fig:16}
\end{figure}

Piecewise continuous or continuous angular distributions are much less common. One of them is shown in Fig.~\ref{fig:16}. As above, a burst of hard radiation occurs approximately at the same instant around 250~ns, near the second peak of the total radiation pulse.

In a recently published paper~\cite{42}, the energy distribution of the gamma-radiation quanta generated in a laboratory atmospheric discharge at a maximum voltage of 950~kV and discharge gap length of 1~m was obtained using calculations and experimental data. According to these results, the energy distribution can be approximated by the bremsstrahlung spectrum of the gamma quanta with a maximum energy of 200--250~keV and an average energy of 52--55~keV.

The obtained experimental data show that the ``average'' energy of the gamma quanta (in the context of the formation of the distributions, which are regular with respect to the angle) exceeds 100~keV, but is smaller than 350~keV, since the needle-like angular distributions are mostly recorded when using the 10-mm-thick Pb filter. As far as the maximum energy is concerned, it can significantly exceed 350~keV according to the results given above.

\subsection{Results on the angular distributions of the radiation in the energy range above 1500~keV}

In the following experiments all on-arc detectors were shielded with an additional 50-mm-thick Pb filter, with tenfold attenuation of the gamma radiation having an energy of 1500~keV (cut-off energy).

\begin{figure}
\resizebox{0.45\textwidth}{!}{\includegraphics*{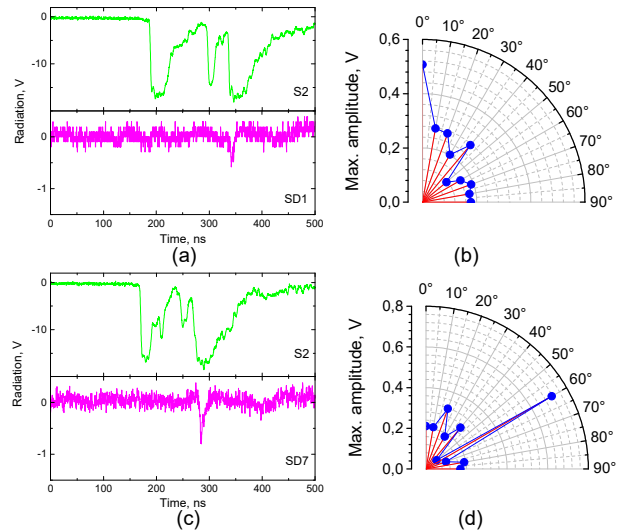}}
\caption{Oscillograms for low-threshold forward detector S2 and oscillograms for on-arc detectors SD1 \textit{a} and SD7 \textit{c}; angular distributions of hard gamma radiation recorded by arc detector system \textit{b, d} for two shots.} \label{fig:17}
\end{figure}

In a series of 54~shots, it was demonstrated that high-energy gamma quanta are present in the radiation accompanying the discharge. The on-arc detectors  recorded extremely anisotropic (needle-like) radiation, but with its amplitude several times smaller than that in Fig.~\ref{fig:14}. Figure~\ref{fig:17} shows the results for two shots. In the first shot, the radiation is directed along the axis, and in the second one, it is directed at an angle to the axis.

\section{Conclusion}

The results of the experiments on the measurement of the angular distributions of the bremsstrahlung of runaway electrons at the initial stage of a laboratory atmospheric high-voltage discharge are presented. The results are in part consistent with the data obtained at other laboratory facilities; however, significant differences are observed in some cases.

The differences are primarily associated with the small-scale (but sometimes very different in amplitude) angular distributions of the bremsstrahlung in the atmospheric discharge, as well as with the energy of the detected gamma quanta. First of all, it was possible to capture these differences by creating a system of scintillation detectors designed as an arc assembly. The angular distributions of the bremsstrahlung of the atmospheric discharge were recorded using the assembly of ten identical scintillation detectors installed uniformly around the circumference of a quarter of a circle. Lead filters having different thicknesses were introduced into the setup to cut off the gamma quanta with energies below the target value specified by the filter thickness. The employment of additional detectors for measuring the radiation in a wide range of energies, including very hard gamma quanta, was also important, as it provided a means to compare the radiation durations and the time of appearance for the quanta of different hardness.

Calculations of the angular distributions of the bremsstrahlung of electrons in air were carried out in the Born approximation for various electron distribution functions normalized to the same number of electrons in the beam. The calculated distribution functions of runaway electrons in the atmospheric discharge were also regarded in the framework of the calculations. The qualitative results turned out to be approximately the same for all considered distribution functions.  On the contrary, the quantitative results were related to the angle at which the radiation maximum appeared. This angle was dependent on the relative intensity of the high-energy electron beam. In order to compare the numerical results and the experimental data, the calculations were carried out in a way to simulate the presence of the filters covering the detectors and absorbing the photons with energies lower than the specified value. With increasing the fraction of high-energy electrons in the considered distribution functions, apart from the narrowing of the angular distribution of radiation, the total energy of the quanta incident on the detectors of the system changes by orders of magnitude. This corresponds to a drastic change in the amplitude of the detector signals and and entails difficulties in their recording. That is, with fixed tuning of the scintillation detectors, which is necessary for quantitative comparison, the amplitudes of the signals can vary by orders of magnitude. Naturally, it is difficult to judge on the temporal structure of the radiation using the considered approach. The difficulties that arise when calculating the angular distribution of the radiation are also related to the fact that the radiation is generated not only by the anode-directed streamers, but by the cathode-directed ones as well. An arc detector system can be used to trace these effects in the experiments.

The obtained experimental data show that, as a rule, one or several flashes of radiation are recorded in each shot, with each of these flashes having its own radiation pattern. In addition, in a considerable number of shots, hard radiation with photon energies comparable to or exceeding the maximum electron energy corresponding to the applied voltage is recorded. In this case, averaging of the experimental data makes no sense, regardless of whether it is done for all shots in any series or for all radiation pulses in each shot. It is unlikely to either provide valuable information due to the inevitable mixing of many factors, or, to result in something that resembles the angular distribution corresponding to one of the calculated ones obtained under well-defined assumptions. In addition, with averaging over a small series of shots, a single shot that produces a large-amplitude signal can result in the suppression of the others.

A series of experiments with the recording of the radiation having various hardness of the quanta (from 10~keV to several hundred keV), show that the radiation in the form of short pulses is present in the discharges. This radiation appears as both single pulses and a sequence of several pulses with different photon energies. The maximum energy of the gamma quanta is not strictly related to the number of the flash. With increasing the cut-off energy of the gamma quanta (increasing the filter thickness), the duration of the harder radiation decreases to values of 4--5~ns. These values are close to the temporal resolution of the photomultipliers used. As a rule, the amplitude of the detector signal drops as well.

In the measurements carried out using 3-mm-thick Al and Pb filters (cut-off energies of the quanta are 20~keV and 100~keV, respectively), a continuous multi-beam radiation pattern with different amplitudes of each beam was recorded in most cases. For several radiation pulses that appeared in this shot, radiation patterns corresponding to the calculated ones were recorded as well.

When the quanta with energies below 350~keV are cut off, the radiation pattern ceases to be continuous with respect to the angle, and takes a needle-like shape with individual detectors of the arc system triggered, including those located at large angles to the axis of the discharge. With an even larger increase in the thickness of the filters, the recorded radiation reveals a pronounced anisotropic (needle-like) character and appears much less frequently (2--3\% of the total number of shots). That is, under these experimental conditions, the maximum electron energy in most of the shots does not exceed $\sim 300$~keV. The electrons of higher energies, including those having the energies comparable to the ``applied voltage'', which provide only a needle-like angular distribution of radiation, are accelerated in different channels. This process possibly occurs through collective fields generated in the created plasma and can play an important role in the development of discharges in a thunderstorm atmosphere.

\begin{acknowledgments}
This work was partially supported by the Russian Foundation for Basic Research, grant No.~17-08-01690.
\end{acknowledgments}

\bibliography{duga}

\end{document}